\documentclass[prc,twocolumn,showpacs,preprintnumbers,amsmath,amssymb,nofootinbib]{revtex4}
\usepackage{graphicx}
\usepackage{amsmath}
\usepackage{bm}
\newcommand{\hfbax}{\sc hfb-ax}

\def\nuc#1#2{\relax\ifmmode{}^{#1}{\protect\text{#2}}\else${}^{#1}$#2\fi}

\begin{document}

\title{ Evolution of surface deformations of weakly-bound nuclei in the continuum}
\author{J.C. Pei}
\affiliation{State Key Laboratory of Nuclear
Physics and Technology, School of Physics, Peking University,  Beijing 100871, China}
\author{Y.N. Zhang}
\affiliation{State Key Laboratory of Nuclear
Physics and Technology, School of Physics, Peking University,  Beijing 100871, China}
\author{F.R. Xu}
\affiliation{State Key Laboratory of Nuclear
Physics and Technology, School of Physics, Peking University,  Beijing 100871, China}

\date{\today}

\begin{abstract}

We study weakly-bound deformed nuclei based on the coordinate-space Skyrme Hartree-Fock-Bogoliubov (HFB) approach,
 in which a large box is employed for treating the continuum and large spatial extensions.
Approaching the limit of the core-halo deformation decoupling, calculations found an exotic ``egg"-like structure consisting of a spherical core plus a prolate halo
 in $^{38}$Ne, in which the near-threshold non-resonant
continuum plays an essential role.
  Generally the halo probability and the decoupling effect in heavy nuclei can be hindered by high
   level densities around Fermi surfaces. However, deformed halos in medium-mass nuclei are possible as
the negative-parity levels are sparse, e.g., in $^{110}$Ge.
The deformation decoupling has also been demonstrated in pairing density distributions.
\end{abstract}


\maketitle

In the exploration for limits of the nuclear landscape,
exotic phenomena such as halo states and the associated soft-mode excitations are expected near the particle drip-lines~\cite{halo-exp}.
The quantum halo state, basically a threshold effect,  is characterized by a tremendous surface diffuseness and  has been observed in weakly-bound nuclei and molecules~\cite{halo}.
In particular, drip-line nuclei are weakly-bound superfluid systems,
in which the pairing induced continuum coupling plays an essential role~\cite{continuum-jacek,continuum,forssen,hagen}.
In deformed weakly-bound nuclei, exotic halo structures could happen due to the core-halo deformation decoupling~\cite{misu,sgzhou}.
Therefore, a crucial issue for theoretical descriptions of weakly-bound nuclei is to be able to self-consistently treat
continuum contributions, deformations and large spatial extensions.

There have been numerous studies of weakly-bound nuclei with assumed spherical shapes and have brought a number of
new insights, for example, the core-halo decoupling~\cite{halo-exp}, the pairing anti-halo effect~\cite{anti-halo}, shell quenching~\cite{jacek94}, and the continuum coupling effect~\cite{continuum-jacek,continuum}. However,  new insights and exotic structures are also expected beyond the spherical shape.
One of the most intriguing structures is the deformed nuclear halo, which can have exotic surface shapes decoupling from cores~\cite{misu,sgzhou}.
In particular, except for $^{14}$Be~\cite{14Be}, the existence of two-neutron deformed halos in heavier nuclei is under question~\cite{nues}.
 Another interesting phenomenon is the predicted deformation difference between neutrons and protons near the neutron drip line,
implying unusual isovector quadrupole collective modes~\cite{misu}. Thirdly, in weakly-bound nuclei, the deformation
of pairing densities (or anomalous density) could be different from particle densities,
which can be studied by pair transfer experiments but have been rarely discussed.

To describe weakly-bound nuclei, especially for medium-mass and heavier nuclei, the HFB approach
 is the theoretical tool of choice by properly taking into the continuum coupling.
For deformed nuclei in the continuum, however, exact HFB solutions with outgoing boundary conditions are very difficult.
The Green function method has made remarkable progress in solving the deformed HFB equation~\cite{oba}, while self-consistent calculations in this way are still missing.
On the other hand, the coordinate-space HFB approach,  referred to as the $\mathcal{L}^2$ discretization method, has been demonstrated to
be able to treat the quasiparticle (qp) resonances and continuum contributions rather accurately~\cite{pei2011}, compared to the exact Gamow HFB solutions~\cite{Michel}.
For weakly-bound deformed nuclei, calculations have to employ a large coordinate-space box that is crucial
for describing the large spatial extension and the continuum discretization~\cite{continuum,anti-halo}. Recently
such very expensive calculations of axial-symmetric deformed nuclei have been realized by performing hybrid parallel calculations on
supercomputing facilities~\cite{pei2012}.

According to systematic calculations of the nuclear landscape, there are several deformed regions close to drip lines~\cite{erler,mario03}.
Experiments towards the neutron drip line have been able to reach $^{40}$Mg and $^{42}$Al~\cite{nscl}, which are predicted to be deformed~\cite{erler,mario03}.
Extensive studies have been done to look for deformed halo structures in light nuclei
experimentally~\cite{halo-rev,14Be,Ne31} and theoretically~\cite{pei06,nakada,sgzhou,lulu}.
For halos in weakly-bound heavy nuclei, theoretical studies are mostly done with spherical nuclear shapes~\cite{meng,Mizutori,schunck,rotival,zhang}.
The studies of deformed heavy halos based on self-consistent HFB calculations are still absent.
Actually the on-going experimental facilities such as FRIB
are expected to provide an unprecedented opportunity for exploring heavy weakly-bound nuclei~\cite{FRIB}. In this context,
we systematically studied the even-even weakly-bound deformed nuclei from light to heavy mass region based on
 the self-consistent coordinate-space Skyrme-HFB approach.
While the odd-$N$ nuclei with much reduced pairing correlations and continuum couplings due to one qp excitation are not involved here.

In this work, the Skyrme-HFB equation is solved by the {\hfbax} code~\cite{Pei08,pei2011,pei2012} within a 2D lattice box, based on B-spline techniques for
axial symmetric deformed nuclei~\cite{teran}. To obtain sufficient accuracy, the adopted 2D box size is 30$\times$30 fm.
 The maxima mesh size is 0.6 fm and the order of B-splines is 12.
This is the first deformed HFB calculations with such a large box size,
 while in literatures the adopted 2D box sizes are usually less than 20 fm. From 20 fm to 30 fm,
 the estimated computing cost will be increased by 40 times~\footnote{The estimated diagonalization cost (the major computing cost) 
 is proportational to $N_R^3$, where $N_R$ denotes the rank of the Hamiltonian matrix.
For a 2D lattice representation,  $N_R$ is is proportational to
 $N_L^2$, where $N_L$ is the one-dimensional grid number. For 20 fm and 30 fm boxes in our approach, $N_L$ are 45 and 62 respectively.}.
For calculations employing large boxes and small lattice spacings,
 the discretized continuum spectra would be very dense and provide good resolutions.
Because the computing cost is extremely high, the hybrid MPI+OpenMP parallel programming is implemented to get
converged results within a reasonable time~\cite{pei2012}.
Calculations were performed in China's top supercomputer Tianhe-1A.
For the particle-hole interaction channel, the SLy4 force~\cite{sly4} is adopted as
it is one of the mostly used parameterizations for neutron rich nuclei.
For the particle-particle channel, the density dependent pairing interactions with the mixed and surface variants
are used~\cite{mix-pairing}. The pairing strengthes are fitted to the neutron gap of $^{120}$Sn.

\begin{figure}[htb]

\centerline{\includegraphics[trim=0cm 0cm 0cm
0cm,width=0.48\textwidth,clip]{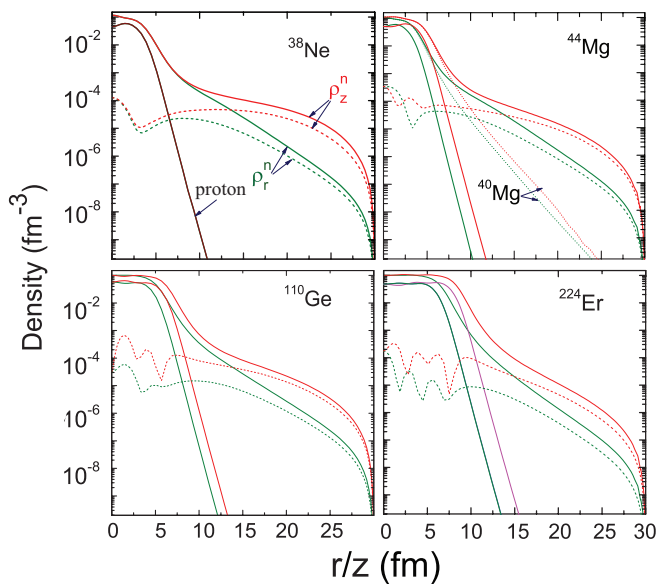}}
\caption{\label{density} (Color online) The density profiles of $^{38}$Ne,
$^{40,44}$Mg, $^{110}$Ge, and $^{224}$Er, which are obtained from calculations with the surface pairing.
The density distributions are displayed along the cylindrical coordinates $z$-axis and $r$-axis, respectively.
The dashed lines show the density distributions contributed by near-threshold qp continuum states with energies below 1.5 MeV.  }
\end{figure}

\begin{figure}[htb]

\centerline{\includegraphics[trim=0cm 0cm 0cm
0cm,width=0.48\textwidth,clip]{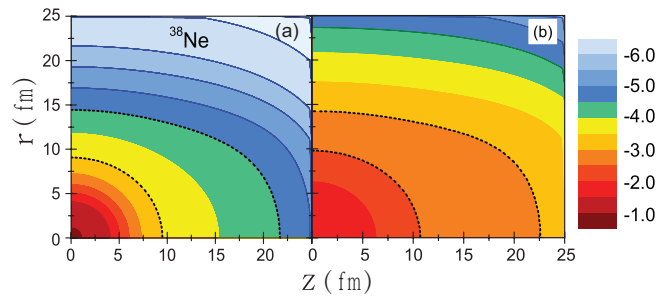}}
\caption{\label{2density} (Color online) The two-dimensional distributions of the neutron density (a) and neutron
pairing density (b) in $^{38}$Ne, which are displayed on the logarithm scale. The shapes of the core and halo
are denoted by dashed lines. Assuming the core has a radius of 9.5 fm, the integrated neutron number inside the core is 26.1, or say, there
are about 2 neutrons in the halo.}
\end{figure}

\begin{figure}[htb]

\centerline{\includegraphics[trim=0cm 0cm 0cm
0cm,width=0.32\textwidth,clip]{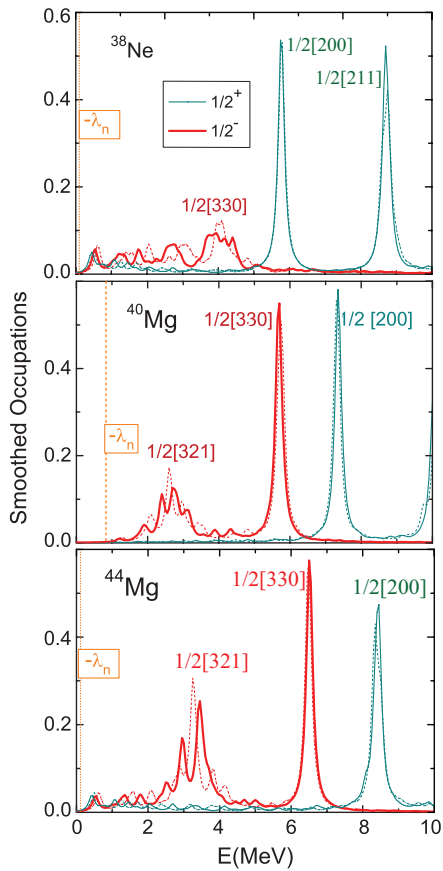}}
\caption{\label{res1} (Color online) The smoothed occupation numbers of $\Omega^{\pi}$=1/2$^{\pm}$ neutron qp states of $^{38}$Ne, $^{40}$Mg and $^{44}$Mg.
The solid lines and dashed lines show the results with box sizes of 30 fm and 27 fm, respectively. The continuum thresholds $-\lambda_n$ are given,
 where $\lambda_n$ is the neutron Fermi surface energy.}
\end{figure}

\begin{figure*}[htb]
\centerline{\includegraphics[trim=0cm 0cm 0cm
0cm,width=0.7\textwidth,clip]{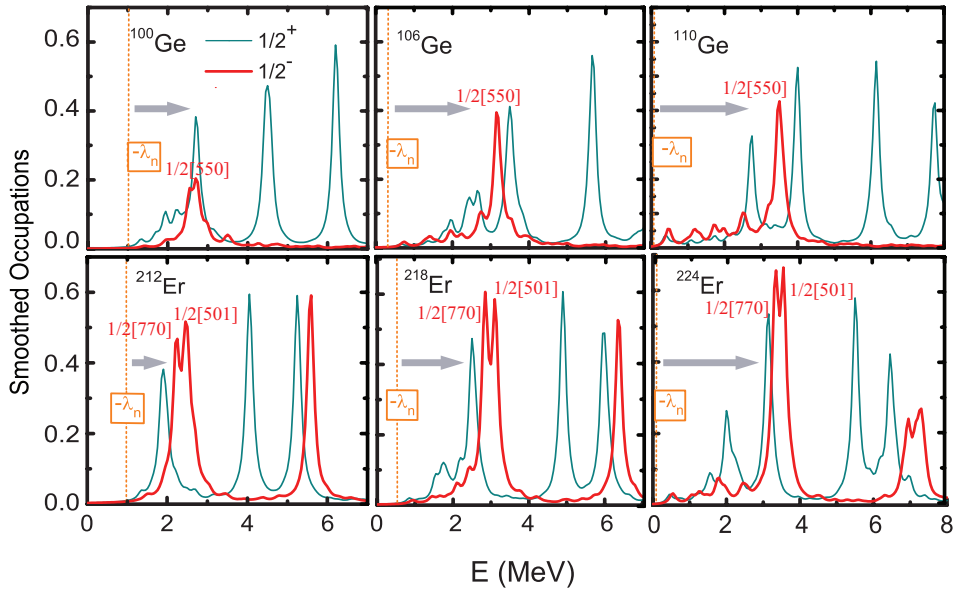}}
\caption{\label{smoothd} (Color online) Similar to Fig.\ref{res1} but for Ge and Er isotopes.
The Nilsson labels are given for near-threshold $\Omega^{\pi}$=$1/2^{-}$ qp states. The arrows in panels are given for
guiding eyes for the evolution of phase space decouplings. See text for details.}
\end{figure*}

We have searched for deformed halo structures along the neutron drip line around three deformed regions:
(I) the Ne and Mg isotopes, (II) the medium-mass region around $Z$=30, and (III) the heavy-mass region around $Z$=60.
In Fig.\ref{density},  the
density profiles of $^{38}$Ne, $^{40, 44}$Mg, $^{110}$Ge, and $^{224}$Er are displayed along
the cylindrical  coordinates $z$-axis (the axis of symmetry)  and $r$-axis (the axis perpendicular to $z$-axis and $r$=$\sqrt{x^2+y^2}$), respectively. Note that the density distributions of axial-symmetric nuclei are represented on the 2-D lattice as $\rho(r_i, z_j)$. The differences between the density profiles $\rho_{z({r=0})}$ and  $\rho_{r({z=0})}$  actually reflect the surface deformations.  The four drip-line nuclei all have
considerable pairing fields as well as continuum contributions.
Results shown in Fig.~\ref{density} are obtained with the surface pairing
since halo characteristics are more significant than with the mixed pairing.
In Fig.~\ref{density}, $^{38}$Ne has the most pronounced  halo structure. Generally halo structures gradually fade away as nuclei towards heavier mass region.
In addition, as nuclei become heavier, one can see that the surface deformations become similar to core deformations,
indicating the weakening of the core-halo deformation decoupling effect. In Fig.\ref{density}, it can also be seen that the contributions of near-threshold qp states (the dashed lines) with energies below 1.5 MeV are mainly responsible for the halo structures and surface deformations. We have to note that the predictions can be dependent on the effective Hamiltonian. In the relativistic Hartree-Bogoliubov calculations, $^{44}$ Mg has a prolate core and a slightly oblate halo while $^{38}$Ne has no pronounced halo~\cite{sgzhou}.

In Fig.\ref{2density}(a), it is very clear to see an exotic ``egg"-like halo structure in $^{38}$Ne, in which the neutron halo distribution
 is well prolately deformed but the core is spherical.
Inside the core, both the proton and neutron density distributions have a spherical shape with a radius of less than 10 fm.
There are about 2 neutrons in the halo by integrating the neutron density beyond the spherical core. The quadrupole deformation of neutrons is $\beta_2^{\rm n}$=0.24, which is completely due to the surface deformation.
 Such an exotic structure, as a manifestation of the core-halo deformation decoupling effect,  has never been obtained in earlier self-consistent calculations, emphasizing the importance
of precisely treating the subtle interplay between the continuum coupling, deformations and surface diffuseness in our approach.
Its neighbor $^{39}$Na could also be a candidate for the ``egg"-like halo structure.
We should note that a possible ``egg"-like structure has also been discussed in $^{16}$C, nevertheless, based on the clustering perspective~\cite{ong,amd}.
In Fig.\ref{2density}(b), the neutron pairing density distribution of $^{38}$Ne also shows the ``egg"-like structure,
reflecting the influence of the deformation decoupling on pairing densities.
The most significant halo structure shown in $^{38}$Ne is consistent with an earlier point that
two-neutron halos could be hindered by deformed cores~\cite{nues}.

To understand the deformed halo mechanism in terms of qp spectra, we display
the qp spectra with smoothed occupation numbers, as shown in Fig.\ref{res1}. It is known that the discretized spectra of qp
resonances can roughly  have the Breit-Wigner shape~\cite{pei2011}. For demonstration, we smoothed the qp
occupations $v_i^2$ with a Lorentz shape function and a smoothing parameter of 0.2 MeV (the smoothing is described in Ref.~\cite{pei2011}).
To distinguish the qp resonances and continuum, we display the qp spectra obtained with two different box sizes of 30 fm and 27 fm, respectively.
It is known that the qp-resonance energies are stationary with different box calculations, while continuum states are not~\cite{jacek84,pei2011}.
In Fig.\ref{res1}, the  smoothed occupations of $\Omega^{\pi}$=1/2$^{\pm}$ neutron qp states of $^{38}$Ne,  $^{40}$Mg  and
$^{44}$Mg are shown, while contributions from high $\Omega$ states are much smaller.
For qp resonances, the related Nilsson labels are given for detailed understandings.
The continuum states with energies below 2 MeV are most likely non-resonant continuum states since their energies are not stationary.
It can be seen that $^{38}$Ne and $^{44}$Mg have obviously larger non-resonant continuum contributions near the threshold compared to $^{40}$Mg.
Thus the halo structures and surface deformations in $^{38}$Ne and  $^{44}$Mg  are mainly due to the near-threshold non-resonant continuum.
According to our calculations, the near-threshold continuum states of both negative- and positive-parity are essential,
although negative-parity continuum contributions are larger.
$^{40}$Mg has the same magic neutron number of $N$=28 as $^{38}$Ne, but it has an abrupt change in qp spectra.
In $^{38}$Ne, the 1/2[330] orbit from $1f_{7/2}$ is only
 partially occupied, and distributed within the {\it spherical core}. $^{40}$Mg has a similar qp structure to $^{44}$Mg except for its much less near-threshold continuum.
$^{38}$Ne has the most significant near-threshold continuum.
One can infer that the ``egg"-like halo structure in $^{38}$Ne can be ascribed
to a considerable $p$-wave scattering into the continuum, i.e., the unbound 1/2[321] orbit (from $2p_{3/2}$). For $^{44}$Mg,  it is due to the $2p_{1/2}$ level embedded in the continuum.
In fact, the qp spectra can contain some unbound single-particle levels coupled to the continuum.
This would be obvious by transforming the HFB spectrum into the BCS spectrum in canonical basis (see Fig.2 in Ref.~\cite{sgzhou}).
In exact HFB solutions or with an even larger box, the unbound single-particle levels should be completely dissolved into the continuum.

In Fig.\ref{smoothd}, the systematics of the smoothed qp spectrum of Ge and Er isotopes are also displayed
to understand the development of deformed halos in heavy nuclei.  From the 1/2[550] orbit in Ge and the 1/2[770] orbit in Er isotopes,
one can see that as isotopes towards the neutron drip
line, the weakly-bound $\Omega^{\pi}=1/2^{\pm}$ qp states move away from the thresholds ($-\lambda_{\rm n}$) collectively.
 Simultaneously,
the near-threshold non-resonant continuum states gradually develop and decouple from cores (or bound states).
 This situation is also seen in $^{40, 44}$Mg as shown in Fig.\ref{res1}.
It seems  that
a large phase space (or gap) between bound states and the thresholds is associated with the development
of the near-threshold continuum and the core-halo decoupling. Indeed, the``egg"-like halo structure in $^{38}$Ne has the largest phase space decoupling.
Furthermore, the phase space decoupling in halos should be model independent,  since a similar situation has also been seen in the relativistic Hartree-Bogoliubov calculation of $^{44}$Mg~\cite{sgzhou}.
In general, such phase spaces, and consequently the halo probabilities and the deformation decouplings,  would be suppressed in heavier nuclei due to higher level densities around Fermi surfaces.
However, in $^{110}$Ge, the negative-parity qp spectrum of $\Omega^{\pi}$=1/2$^{-}$ is sparse around the neutron Fermi surface.
In $^{110}$Ge, the near-threshold 1/2$^{-}[550]$ orbit (belonging to $1h_{11/2}$)  is the only negative-parity level from $N$=50 to $N$=82, leading to a large phase space
decoupling. The deformed halo in $^{110}$Ge is mainly due to the $2f_{7/2}$ orbit 1/2[541] coupled with the continuum.
It can be seen that $^{224}$Er has less near-threshold continuum contribution than $^{110}$Ge although they have similar phase spaces in the $1/2^{-}$ spectra, because the high-$j$ orbit 1/2[761] above the threshold (compared to 1/2[541] in $^{110}$Ge)
experiences much less continuum coupling effects~\cite{continuum}.
 No considerable two-neutron halos and deformation decoupling effects
are obtained from Nd to Er nuclei.

\begin{figure}[t]

\centerline{\includegraphics[trim=0cm 0cm 0cm
0cm, width=0.45\textwidth,clip]{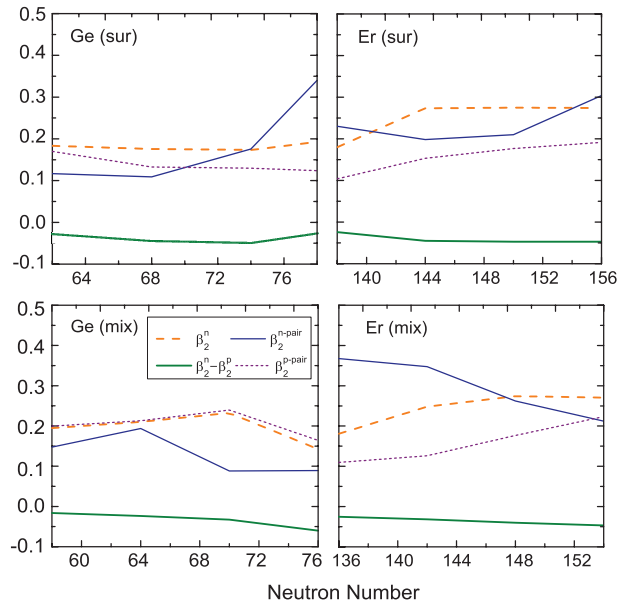}}
\caption{\label{deformations} (Color online) Quadrupole deformations of neutrons ($\beta_2^{\rm n}$, dashed lines), neutron pairing densities ($\beta_2^{\rm n-pair}$, solid lines),
proton pairing densities ($\beta_2^{\rm p-pair}$, short-dashed lines),
 and isovector deformations ($\beta_2^{\rm n}-\beta_2^{\rm p}$, thick-solid lines) in Ge and Er isotopes. Results are obtained with
 the surface pairing (the upper panel) and the mixed pairing (the lower panel) interactions, respectively. }
\end{figure}

Besides the core-halo deformation decoupling, the neutron-proton deformation difference, i.e., the isovector deformation, is also
very interesting.
It is known that generally nuclei close to the neutron drip line can have considerable negative isovector deformations, implying unusual
isovector quadrupole collective modes~\cite{erler,misu}. In this work, we calculated
the quadrupole deformations $\beta_2$ of both particle-density and pairing-density distributions using the formula:
$\beta_2 = ({4\pi}/5){ <r^2Y_{20}>}/{<r^2>}$~\cite{erler}.
The isovector deformations, deformations of neutrons and pairing densities are displayed in Fig.\ref{deformations}.
One can see that similar negative isovector deformations are obtained in calculations with both the mixed and surface pairing
interactions.
With the mixed pairing, the absolute isovector deformations tend to increase towards the neutron drip line.
With the surface pairing, the neutron drip-lines of Ge and Er have two more neutrons in addition to the mixed pairing.
One can see that deformations of neutron pairing density distributions obtained in surface pairing calculations increase
near the neutron drip line.
The obtained isovector deformations of Ge and Er isotopes are almost the same as the HO-basis HFB calculations~\cite{hfbtho}.
However, $^{38}$Ne and $^{44}$Mg are slightly unbound in the HO-basis approach.
With the surface pairing for $^{38}$Ne and $^{44}$Mg, the isovector quadrupole deformations
 are 0.24 and $-$0.09, respectively.
The corresponding quadrupole deformations of neutrons (and neutron pair densities)
 are 0.24 (0.48) and 0.25 (0.36), respectively. We can see that $^{38}$Ne (shown in Fig.\ref{2density}), $^{44}$Mg, and $^{110}$Ge,
 all have much larger deformations of neutron-pairing densities than neutron densities, due to remarkable non-resonant continuum contributions.
 It has to be noted that in $^{38}$Ne the isovector deformation has an abnormal large positive value because of the ``egg"-like halo structure.
It can also be seen that deformations of neutron pairing densities are generally different from that of neutron densities
and are sensitive to the effective pairing interaction.

In summary, we have studied the weakly-bound deformed nuclei
based on the self-consistent coordinate-space Skyrme-HFB approach, to look for new insights
by treating
the continuum within a large box.
We found an exotic ``egg''-like halo structure in $^{38}$Ne for the first time with a spherical core plus a prolate halo, for which
the near-threshold non-resonant continuum is essential. In addition, the ``egg''-like halo structure has a large positive isovector deformation,
implying unusual isovector quadrupole modes and further
experimental observations (or its neighbor $^{39}$Na) will be very important.
Compared to light nuclei, the halo probability and the core-halo deformation decoupling effect in heavy nuclei could be hindered by high
level densities around Fermi surfaces. However, $^{110}$Ge has a sparse negative-parity spectrum and shows a deformed halo.
The deformation decoupling has also been shown in the pairing density distributions.  Nevertheless, the deformations
of pairing density distributions are different from particle densities and are sensitive to the effective pairing Hamiltonian.

\section*{Acknowledgments}
We thank W. Nazarewicz, R. Wyss, A.T. Kruppa, M. Kortelainen and D.Y. Pang for useful discussions.
 This work was supported by the
National Key Basic Research Program of China under Grant 2013CB834400,
and the National Natural Science Foundation of China under Grant No.11235001.
We also acknowledge that computations in this work were performed in the Tianhe-1A supercomputer
 located in the Chinese National Supercomputer Center in Tianjin.


\begin{thebibliography}{999}

\bibitem{halo-exp}
I. Tanihata, J. Phys. G 22, 157(1996).

\bibitem{halo}
A.S. Jensen, K. Riisager, D.V. Fedorov and E. Garrido, Rev. Mod. Phys. 76,  215(2004).

\bibitem{continuum-jacek}
J. Dobaczewski, W. Nazarewicz, T. R. Werner, J. F. Berger, C. R. Chinn, and J. Decharg\'{e},
Phys. Rev. C {\bf 53}, 2809(1996).

\bibitem{continuum}
M. Yamagami, Phys. Rev. C {\bf 72}, 064308 (2005).

\bibitem{forssen}
C. Forssen, G. Hagen, M. Hjorth-Jensen, W. Nazarewicz, J. Rotureau, Physica Scripta T152, 014022 (2013).

\bibitem{hagen}
G. Hagen, T. Papenbrock, and M. Hjorth-Jensen, Phys. Rev. Lett. 104, 182501(2010).

\bibitem{misu}
 T. Misu, W. Nazarewicz, S. Aberg, Nucl. Phys. A 614, 44 (1997).

\bibitem{sgzhou}
S.G. Zhou, J. Meng, P. Ring, and E.G. Zhao, Phys. Rev. C 82, 011301(R)(2010).

\bibitem{anti-halo}
K. Bennaceur, J. Dobaczewski, M. Ploszajczak, Phys. Lett. B 496, 154(2000).

\bibitem{jacek94}
J. Dobaczewski, I. Hamamoto, W. Nazarewicz, and J.A. Sheikh, Phys. Rev. Lett. 72, 981 (1994)

\bibitem{14Be}
M. Labiche {\it{et al.}}, Phys. Rev. Lett. 86, 600 (2001).

\bibitem{nues}
F.M. Nunes, Nucl. Phys. A 757, 349(2005).

\bibitem{oba}
{H. Oba and M. Matsuo, Phys. Rev. C {\bf 80}, 024301 (2009)}.

\bibitem{pei2011}
J.C. Pei, A.T. Kruppa, and W. Nazarewicz, Phys. Rev. C 84, 024311 (2011).

\bibitem{Michel}
{ N. Michel, K. Matsuyanagi, and M. Stoitsov, Phys. Rev. C {\bf 78}, 044319
  (2008).}

\bibitem{pei2012}
J.C. Pei, G.I. Fann, R.J. Harrison, W. Nazarewicz, J. Hill, D. Galindo, J. Jia,  J. Phys. Conf. Ser. 402, 012035(2012).

\bibitem{erler}
J. Erler, N. Birge, M. Kortelainen, W. Nazarewicz, E. Olsen, A.M. Perhac, M. Stoitsov,
Nature 486, 509 (2012).

\bibitem{mario03}
M. V. Stoitsov, J. Dobaczewski, W. Nazarewicz, S. Pittel, and D. J. Dean,
Phys. Rev. C 68, 054312 (2003).

\bibitem{nscl}
T. Baumann et al., Nature 449, 1022(2007).

\bibitem{halo-rev}
B. Jonson, Phys. Rept. 389, 1(2004).

\bibitem{Ne31}
T. Nakamura et al., Phys. Rev. Lett. 103, 262501(2009).

\bibitem{pei06}
J.C. Pei, F.R. Xu, P.D. Stevenson, Nucl. Phys. A 765, 29 (2006).

\bibitem{nakada}
H. Nakada, Nucl. Phys. A 808, 47 (2008).

\bibitem{lulu}
Lulu Li, J. Meng, P. Ring, E.G. Zhao, and S.G. Zhou,
Phys. Rev. C 85, 024312(2012).


\bibitem{meng}
J. Meng and P. Ring, Phys. Rev. Lett. 80, 460(1998).

\bibitem{Mizutori}
S. Mizutori, J. Dobaczewski, G. A. Lalazissis, W. Nazarewicz, and P.-G. Reinhard,
Phys. Rev. C 61, 044326(2000).

\bibitem{schunck}
N. Schunck and J.L. Egido, Phys. Rev. C 78, 064305(2008).

\bibitem{rotival}
V. Rotival and T. Duguet, Phys. Rev.  C 79, 054308(2009).


\bibitem{zhang}
Y. Zhang, M. Matsuo, J. Meng, Phys. Rev. C 86, 054318(2012).


\bibitem{FRIB}
Refer to B. Sherrill's talk in the ``Nuclear Structure 2012" conference (http://ns12.anl.gov).

\bibitem{Pei08}
{J. C. Pei, M. V. Stoitsov, G. I. Fann, W. Nazarewicz, N. Schunck, and F. R.
  Xu, Phys. Rev. {\bf C78}, 064306 (2008).}

\bibitem{teran}
{E. Ter\'{a}n, V.E. Oberacker, and A.S. Umar, Phys. Rev. C {\bf 67}, 064314
  (2003)}.

\bibitem{sly4}
{ E. Chabanat, P. Bonche, P. Haensel, J. Meyer, and R. Schaeffer, Nucl. Phys. A
  {\bf 635}, 231 (1998)}.

\bibitem{mix-pairing}
{J. Dobaczewski, W. Nazarewicz, and M.V. Stoitsov, Eur. Phys. J. A {\bf 15}, 21
  (2002)}.


\bibitem{ong}
H.J. Ong {\it{et al.}}, Phys. Rev C 73, 024610 (2006).

\bibitem{amd}
Y. Kanada-En¡¯yo, Phys. Rev C 71, 014310 (2005).

\bibitem{jacek84}
J. Dobaczewski, H. Flocard and J. Treiner, Nucl. Phys. A 422, 103 (1984).

\bibitem{hfbtho}
M.V. Stoitsov, J. Dobaczewski, W. Nazarewicz, and P. Ring, Comput. Phys. Commun. 167, 43 (2005).



\end{thebibliography}
\end{document}